# Understanding Leakage Currents through $Al_2O_3$ on $SrTiO_3$

Dror Miron,[1] Igor Krylov,[1] Maria Baskin,[1] Eilam Yalon[1] and Lior Kornblum[1,a)]

[1]*The Andrew & Erna Viterbi Department of Electrical Engineering, Technion - Israel Institute of Technology, Haifa 32000, Israel*

Leakage currents through insulators received continuous attention for decades, owing to their importance for a wide range of technologies, and interest in their fundamental mechanisms. This work investigates the leakage currents through atomic layer deposited (ALD) $Al_2O_3$, grown on $SrTiO_3$. This combination is not only a key building block of oxide electronics, but also a clean system for studying the leakage mechanisms without interfacial layers that form on most of the conventional bottom electrodes. We show how tiny differences in the deposition process can have a dramatic effect on the leakage behavior. Detailed analysis of the leakage behavior rules out Fowler-Nordheim tunneling (FNT) and thermionic emission, and leaves the trap-related mechanisms of trap-assisted tunneling (TAT) and Poole-Frenkel as the likely mechanisms. After annealing the sample in air, the currents are reduced, which is ascribed to transition from trap-based mechanism to FNT, due to the elimination of the traps. The dramatic role of the assumptions regarding the flat-band voltage used for analysis is critically discussed, and the sensitivity of the extracted parameters on this magnitude is quantitatively described. We show that field effect devices based on structures similar to those described here, should be able to modulate $>10^{13}$ $cm^{-2}$ electrons. These results provide general guidelines for reducing and analyzing leakage currents in insulators, and highlight some of the possible approaches and pitfalls in their analysis.

---

a) Author to whom correspondence should be addressed. Electronic mail: liork@ee.technion.ac.il

## I. Introduction

Alumina ($Al_2O_3$) grown by atomic layer deposition (ALD) is a widespread insulating oxide. The motivation for studying this material ranges from understanding various fundamental physical aspects of $Al_2O_3$[1,2] to its considerable potential for applications in electronics, optics and many other fields. Some two decades ago, ALD $Al_2O_3$ received considerable attention as a potential gate insulator for Si technology[3], owing to its large band gap. Despite the preliminary interest, $Al_2O_3$ was eventually sidelined by Hf-based oxides as the high-k gate insulator for ultra-scaled Si logic devices[4]. Nonetheless, $Al_2O_3$ has found other uses in Si technology, such as in ultra-thin layers for effective work-function adjustment[5,6]. Beyond Si technologies, ALD $Al_2O_3$ emerged as the best passivation layer for Ge[7], high-Ge content SiGe devices[8] and with III-V based devices as well[9]. Other back-end microelectronics applications of ALD-$Al_2O_3$ include metal-insulator-metal (MIM) capacitors for resistive switching random-access memory (RRAM) devices[10], antifuse devices[11] and others[12].

More recently, the wide bandgap semiconductor $\beta$-$Ga_2O_3$ has emerged as a promising candidate for power devices[13]. In such roles, the ability to apply high fields is critical for the performance of power field effect devices, and here the high bandgap of $Al_2O_3$ is attractive in reducing the gate leakage currents[14,15].

Another potential application of ALD-$Al_2O_3$ is in oxide electronics, a field greatly invigorated by the discovery of a 2D electron gas (2DEG) at the interface between some insulating oxides[16,17]. One of the promising applications is an oxide field effect transistor (FET) which utilizes the 2DEG as a confined electron channel[18]. Such devices were demonstrated by epitaxially growing $LaAlO_3$ (LAO) on top of single crystal $SrTiO_3$ (STO) substrates[19,20,21]. The subsequent discovery of oxide 2DEG based on amorphous oxides grown on STO[22] has quickly paved the way to the application of ALD-$Al_2O_3$ for this purpose as well[23,24]. This concept has been extended by the replacement of STO crystals by thin $TiO_2$ layers, also grown by ALD[24]. This considerably increases the scalability of oxide electronics, by circumventing the use of single crystalline STO substrates[25], available in limited sizes. Similar $Al_2O_3$/STO and $Al_2O_3$/$TiO_2$ structures have also been suggested as selectors for memristor crossbar arrays,[26] for gas sensors[27] and for spintronic devices[28,29].

The wide band gap of STO of 3.2 eV [23] makes leakage reduction even more challenging, owing to the relatively low possible barriers with the insulator. We previously addressed this issue spectroscopically, by investigating the band alignment at $Al_2O_3$/STO interfaces, and reported barriers of 2.0 ± 0.3 eV and 1.4 ± 0.2 eV for electrons and holes, respectively[30].

One possible explanation to the discrepancy between the large barriers determined by spectroscopy and the high leakage currents is electron traps or states that may exist inside the $Al_2O_3$ band gap, that are typically attributed to oxygen vacancies[31,32]. Therefore, large bandgap and large barriers are insufficient for mitigating gate leakage.

$Al_2O_3$ growth is the easiest and most widespread ALD process, and it can be robustly performed over wide range of temperatures and other process conditions[33] with excellent results. While many applications are relatively insensitive to the growth conditions, thin gate insulators can be extremely dependent of growth parameters. Otherwise-excellent $Al_2O_3$ films may exhibit high leakage currents, poor reproducibility, reliability problems and other issues.

Considering their importance, leakage currents through ALD $Al_2O_3$ have been an integral part of its development since its early days on Si[3]. Understanding the fundamental properties of the leakage currents can be done by studying leakage through ALD-$Al_2O_3$ grown on a semiconductor, on metal-coated substrates or on conductive oxides. However, interfacial layers are typically formed at oxide/semiconductor interfaces, which complicate the analysis of the leakage through the $Al_2O_3$ layer[3]. Many metals form a native oxide surface layer, and when used as the back electrode this layer adds an additional insulator in series; metals that don't have surface oxides, such as Pt, are typically problematic for nucleation of many ALD oxides due to their surface chemistry, which can result in lower quality films.

Conductive oxides are therefore more suitable substrates for studying leakage through ALD-$Al_2O_3$. With $β$-$Ga_2O_3$ devices, the oxide substrate is already part of the device, and this issue has been addressed by several works; for example, Hung et al.[34] and Bhuiyan et al.[15], who both reported trap-assisted tunneling as the dominant $Al_2O_3$ leakage mechanism and extracted a trap energy of 1.1 eV below the conduction band. Another conductive oxide back electrode, indium tin oxide (ITO), has been employed by Spahr et al.[35], who reported a thorough investigation of the leakage currents through low-temperature ALD-$Al_2O_3$, grown at 80 °C. While low temperature processes are crucially important for some applications, their resulting stoichiometry can be less ideal than films grown at 200-300 °C, which is important for leakage reduction.

In this work we address the leakage currents of ALD-$Al_2O_3$ grown on conductive STO substrates. We briefly demonstrate that a default ALD recipe is far from ideal for this task, and by comparison to a more optimized process we obtain further insight into the conduction process. The motivation for this study is twofold: to evaluate ALD-$Al_2O_3$ and its limits for STO-based oxide electronics, and by employing STO as a conductive back-electrode we aim to understand the leakage mechanisms through $Al_2O_3$, a question that is applicable to many technologies beyond oxide electronics.

**II. Experimental**

(001) 0.01%(wt) Nb-doped STO (Nb:STO) crystals (CrysTec GmbH) were $TiO_2$ terminated using the "extended Arkansas" method[36]. This process started with solvent sonication cleaning, followed by a 3:1 HCl-$HNO_3$ treatment. A two-step anneal was performed, starting with 1000 °C for 1 hour in air and followed by 650 °C for 30 minutes in flowing $O_2$. Nominally 10 nm thick amorphous $Al_2O_3$ layer was grown by ALD (Ultratech/Cambridge Nanotech Fiji G2) using trimethyl-aluminum (TMA) and water as the precursors. Two recipes were used and compared: Recipe A, the manufacturer's default for $Al_2O_3$ was performed at a substrate temperature of 300 °C and Recipe B was optimized by extending the water pulse by a factor of 5, the TMA purging pulse by ×2.5 and the water purging pulse by ×1.25, at a substrate temperature of 280 °C. Film thickness of 10±0.5 nm, was measured by x-ray reflectivity (XRR, acquired using a Rigaku SmartLab and analyzed with GlobalFit 2.0). 50 nm Pt pads were deposited through a shadow mask using e-beam evaporation and back contact was prepared by e-beam evaporation of 300 nm blanket Al on the back of the wafer. Current density-voltage (JV) and capacitance-voltage (CV) characteristics were measured using a Keithely 2450 source meter instrument, and a Keysight E4980A precision LCR meter, respectively, in a shielded light-sealed box with a home-built heating stage. After measuring the sample, it was annealed in air in a tube furnace for 30 min at 500 °C (measured on the outer tube surface).

### III. Results and Discussion

The Pt/Al$_2$O$_3$/Nb:STO structures are treated as metal insulator semiconductor (MIS) capacitors, where in the general case, the applied voltage on the gate (V$_g$) can be expressed by the following[37]

$$V_g = V_{FB} + V_{ox} + \psi_s \qquad (1)$$

where $V_{FB}$ is the flat-band voltage, V$_{ox}$ is the voltage drop across the oxide and $\psi_s$ is the band bending, or surface potential of Nb:STO. Since Moon et al.[38] calculated a significant $\psi_s$ in 0.7% doped samples, here CV measurements were conducted at a frequency range of 5-800 kHz to assess the possibility of depletion in the Nb:STO. Measurements were conducted in the voltage ranges of -4 V to 4 V where leakage is undetectable (Figure 2a), in order to ensure our accurate interpretation of the capacitance. Figure 1 shows little voltage or frequency dependence of the capacitance. No significant capacitance reduction is observed with voltage, indicating negligible depletion of the Nb:STO in the measured voltage range. We therefore conclude that $\psi_s$ is insignificant in Eq. (1), and thus the Nb:STO Fermi level remains near the conduction band edge of the highly-doped STO. The flat-band voltage is taken as $V_{FB} = \phi_{Pt} - \phi_{STO} = 1.45\ V$ where $\phi_{Pt}$ is the Pt effective work function that was measured as 5.35 V on Al$_2$O$_3$[39], and $\phi_{STO}$ is approximated as the STO electron affinity of 3.9 V[40]. Additional possible contributions to V$_{FB}$ are neglected at this point, an assumption to be revisited later, and since $\psi_s$ is small, the electric field in the oxide, $E$, is taken as

$$E = V_{ox}/d = (V_g - V_{FB})/d \qquad (2)$$

where d is the oxide thickness. In the absence of Nb:STO depletion, the Al$_2$O$_3$ permittivity can be extracted directly from the CV plot using the parallel plate capacitor expression C$_{ox}$/A=$\varepsilon_0\varepsilon_r$/t$_{ox}$, where $\varepsilon_0$ is the vacuum permittivity, $\varepsilon_r$ is the relative permittivity of Al$_2$O$_3$ and A is the pad area.

Capacitance was extracted from the complex impedance using the $C_s$-$R_s$ model[41] (Fig. 1, inset), where $C_s$ is the series capacitance and $R_s$ the series resistance, acquired at low and high frequencies, respectively. At accumulation (+4V), the extracted $C_s$ at 5 kHz and $R_s$ at 800 kHz before thermal anneal are 0.68 µF/cm² and 32 Ω, respectively, while after thermal anneal the extracted $C_s$ at 5 kHz and $R_s$ at 800 kHz are 0.74 µF/cm² and 94 Ω, respectively (slight oxidation of the surface of the Al back contact may account for this small $R_s$ increase). Relative permittivity of 7.7 and 8.4 was extracted before and after thermal anneal, respectively, in agreement with previous reports[40,42]. The small increase in the capacitance can be ascribed to slight densification, which could increase the permittivity[43] and reduce the thickness.

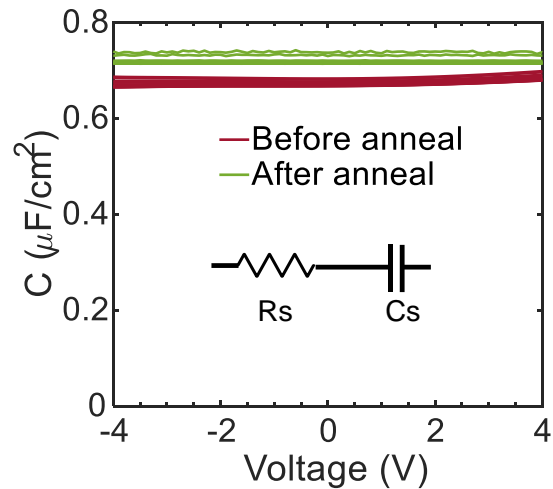

**Figure 1.** CV measurements for 0.01%(wt) Nb:STO grown with recipe B using $C_s$-$R_s$ model (inset) in the frequency range of 5-800 kHz, before and after thermal anneal.

We start the leakage analysis by comparing ALD Recipe A (instrument default) with Recipe B (an optimized recipe). The JV behavior (Figure 2a) shows that the default Recipe A exhibits considerably higher leakage currents at lower voltages. At the most extreme cases, Recipe A has over 2 orders of magnitude higher leakage currents at the same voltage compared to the optimized Recipe B (e.g. around 4 V). Altogether, our optimized recipe yields undetectable leakage (<10⁻⁸ A/cm²) at fields of under ±4 MV/cm, and no signs of breakdown below 5 MV/cm in the devices tested here. In comparison, Moon et al. demonstrated 4 nm Al₂O₃/Nb:STO heterojunction and observed leakage of above 10⁻³ A·cm⁻² at an approximated electric field of 2.5 MV·cm⁻¹ (Ref. [38]); however, at these ultrathin dimensions, additional mechanisms may come into play. These results illustrate how seemingly-identical Al$_2$O$_3$ layers grown under similar conditions can vary wildly in their performance as gate insulators, as a result of minute process details.

From this point onward all analysis is done on Recipe B samples. Beyond the measurement voltage range, devices were found to be prone to irreversible damage. The relevant region for positive bias leakage was analyzed at varying temperatures. Figure 2b exhibits distinct temperature dependence and a current increase of ×2 was measured by raising the temperature from 22 to 70 °C. The sample was measured once again after anneal (500 °C for 30 min in air), and showed two main noticeable differences: for any given voltage value the leakage is noticeably smaller, and more importantly, the temperature dependence was diminished, a fact that would be addressed later.

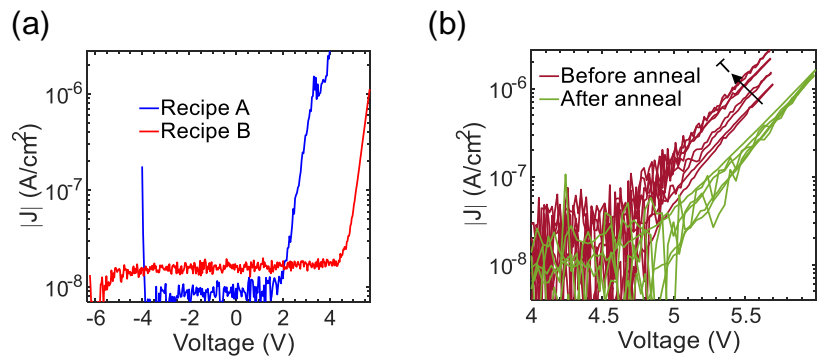

**Figure 2.** (a) JV comparison between recipe A and recipe B. The measured noise floor of up to 4.5 pA at 22 °C corresponds to 2·10$^{-8}$ A·cm$^{-2}$. (b) Varying-temperature JV taken at the positive bias leakage region of Recipe B, before and after anneal. Before anneal the temperature ranges from 22 to 70 °C and after anneal the range is 22 to 50 °C.

Four possible leakage mechanism were considered: Fowler-Nordheim tunneling (FNT), trap-assisted tunneling (TAT), Poole-Frenkel (PF) emission and Schottky-Richardson thermionic emission (TE). The temperature dependence contradicts the FNT mechanism, where no temperature dependence is expected in its most simplified picture. On the other hand, PF and TE and TAT models qualitatively agree with the measured temperature trend.

The positive bias of the JV curve (5-5.7 V) was fitted by the abovementioned mechanisms (Figure 3). The relationship between the current density and the electric field in PF emission model is given by[44]:

$$J \propto E \exp\left[-q\left(\phi_t - \sqrt{qE/(\pi\varepsilon_0\varepsilon_r)}\right)/kT\right]. \quad (3)$$

Where E is electric field in the insulator, q is the electron charge, $\phi_t$ is the trap energy level below the $Al_2O_3$ conduction band, $\varepsilon_r$ is the $Al_2O_3$ (high-frequency) relative permittivity and k is the Boltzman constant. The averaged relative permittivity is extracted from the slope of the linear fit of PF (Figure. 3b), determined to be $3.4 \pm 0.5$, which is somewhat higher than the extracted value from optical measurements[45] of 2.2-2.6[46,47]. Furthermore, other works suggested PF has a negligible probability to be a dominant mechanism[48].

However, The initial use of a value of 1.45 V as the flat-band voltage is an approximation of the effective work functions difference, neglecting all other potential parasitics such as fixed oxide charges that are likely to affect the flat-band voltage[44]. Therefore, a range of flat-band voltages was considered to examine the effect of this elusive magnitude on the parameters extracted from each model. Analyzing the PF conduction mechanism with a broader range of flat-band voltages, yields reasonable flat-band range of voltages of -0.3 to -1.3 V, in which PF emission is possible (Figure 3c) and thus cannot be eliminated.

The relationship between the current density and the electric field in TE model is given by[44]:

$$J = A^* T^2 \exp\left[-q\left(\phi_B - \sqrt{qE/(4\pi\varepsilon_0\varepsilon_r)}\right)/kT\right] \quad (4)$$

Where $\phi_B$ is energy barrier between the conduction band edges of Nb:STO and $Al_2O_3$ and A* is the effective Richardson constant. An average energy barrier of $1.7 \pm 0.1$ eV was extracted from the intercept of the linear fit of TE (Figure 3d), which is in reasonable agreement with the barrier of $2.3 \pm 0.3$ eV, obtained by spectroscopy[30]. However, the averaged relative permittivity of $0.8 \pm 0.1$, extracted from the slope, does not agree with the reported value of ~2.4[46]. While good agreement can be obtained for the barrier $\phi_B$, no flat-band voltage in the (arbitrary) range examined could provide a physically-relevant value for the permittivity (Figure 3f), which further validates the exclusion of the TE mechanism.

Next, we consider the validity of TAT mechanism, where the relationship between the current density and the electric field, according to Fleischer et. al,[49] can be simplified by the following expression:

$$J = 2C_t N_t q\phi_t \cdot (3E)^{-1} \exp\left(-A\phi_t^{-3/2} E^{-1}\right) \qquad (5)$$

where $C_t$ is a slowly varying function of electron energy,[50] $N_t$ the trap density, $A = 4\sqrt{2qm^*} \cdot (3\hbar)^{-1}$, m* is the electron's effective mass in $Al_2O_3$ and $\hbar$ is the reduced Planck constant. The value used for m* is $0.23m_0$ (Ref. 3). This specific TAT model was chosen for its simplicity and its ability to analyze the physical parameters, but we note that TAT modeling has many other variants.

By plotting $\ln(JE)$ as a function of $E^{-1}$ we extract an average trap energy level of 1.6 eV below the conduction band edge of $Al_2O_3$ (Figure 3h). Theoretical analyses of oxygen vacancies in $Al_2O_3$[51,31,32] predict typical energy levels of 2 eV below the conduction band, in some agreement with this experimental observation. After extracting these parameters, the accuracy of this procedure is examined by simulating the TAT integral $J = \int_0^{X_1} qC_t N_t [P_1 P_2 / (P_1 + P_2)] dx$ where $X_1 = (V-\phi_t)/E$ and $P_1$ and $P_2$ are the tunneling probabilities[50]. This simulation is presented in Figure 3h and implies that this simplified model is a good approximation for the TAT integral. TAT provides possible values for the trap levels that vary slowly with $V_{FB}$, and none can be ruled out. The trap density $N_t$, is seen to vary wildly with the assumed $V_{FB}$ (Figure 3i); changes of ~1 V in the former result in two orders of magnitude difference in the latter. We therefore conclude that extracting the trap density from this model is unreliable in the absence of precise knowledge of the flat-band voltage, and that TAT remains a likely candidate. While the basic TAT models are temperature-independent, some temperature dependence has been observed with TAT[38] and modeled by Yu et al[52]. who accounted for the Fermi-Dirac distribution at the injecting electrode; however, simulation using this model could not yield a satisfying fit for all data as well. We've therefore ruled out TE, and cannot rule out TAT and PF at this point.

Last, we examine FNT, where the relationship between the current density and the electric field is given by the following expression[44]:

$$J \propto E^2 \exp\left[-4\sqrt{2qm^*}\phi_B^{3/2} \cdot (3\hbar E)^{-1}\right] \quad (6)$$

An FNT linear fit is shown in Figure 3k. An average barrier height of 1.4 V is extracted from the linear slopes of $\ln(J \cdot E^{-2})$ versus $E^{-1}$ plots, which does not agree with the spectroscopically-measured $2.3 \pm 0.3$ V[30]. The flat-band analysis (Figure 3l) yields good agreement to the expected barrier value of 2.3 eV at low flat-band voltages. While this may appear a success, the increase of the currents with the temperature does not agree with the basic FNT model. Modifications to FNT can incorporate temperature dependence[53], but FNT should produce the lowest currents from the all the mechanisms discussed above; the observation of lower leakage currents after anneal therefore strongly indicates a trap based conduction mechanism (either PF or TAT) as the leakage mechanism for the unannealed sample. As a result of the large energy barrier between STO and $Al_2O_3$, it can be safely assumed that the first step of the dominant conduction mechanism, whatever it is, is tunneling electrons from STO to traps inside $Al_2O_3$. Subsequently, the electrons either gain enough thermal energy to escape out to the conduction band (PF) as modeled by Jeong et al.[54] in their so called 'tunnel assisted PF' (TAPF), or tunnel to the other electrode, as was indicated by Yu et al.[52].

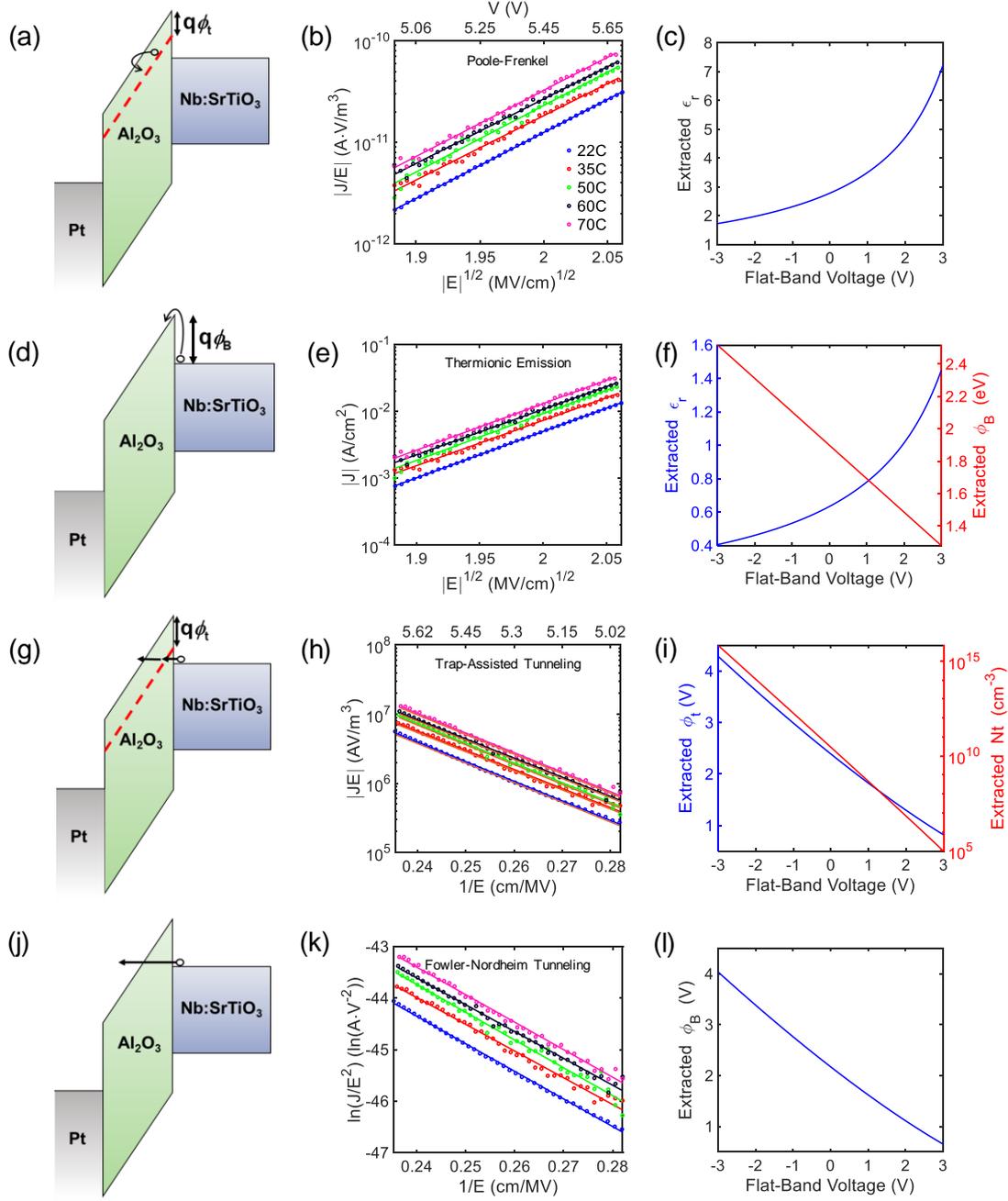

**Figure 3.** Conduction mechanisms analysis at positive bias. (a-c) Poole-Frenkel (PF), (d-f) Thermionic emission (TE), (g-i) Trap-assisted tunneling (TAT) and (j-l) Fowler-Nordheim tunneling (FNT). The first panel of each row shows a schematic of the conduction mechanism. The second panel presents the temperature-dependent fits to the model represented in that row, assuming $V_{FB}$ = 1.45 V. Symbols represent measured data points and lines represent their linear fits. An additional orange line in panel (h) represents the simulated TAT curves (which coincides with the fits, see text). The third panel of each row illustrates the influence of the flat-band voltage assumption on the parameters extracted from that model.

Since the slopes of the JV curves before and after anneal are similar, repeating the analysis of Figure 3 on the annealed samples produces nearly identical parameters. However, the disappearance of the temperature dependence following anneal, and the overall leakage reduction, favor FNT. Since our interpretation of TAT or PF in the unannealed samples ascribed the enhanced electron conductivity to the presence of oxygen vacancies,[38,51] it is quite likely that anneal in air oxidized the sample enough to decrease the trap density and hence minimize the contribution of the traps, leaving FNT as the dominant current mechanism.

For oxide electronics, for example field-effect transistors (FETs) with 2DEG channels, depleting the 2DEG is necessary for closing this "normally-on" device[55]. We therefore estimate the charge modulation $\Delta Q = C_{Al2O3}/\Delta V$ for the unannealed structures. We obtain a modulation of $\sim 4 \times 10^{12}$ electrons/cm$^2$ per 1 V on the gate(0.45 MV/cm using $V_{FB}$ = 1.45 V), or a modulation of $\sim 1.7 \times 10^{13}$ electrons/cm$^2$ at gate voltages of up to |4| V (2.55 MV/cm), before the onset of detectable leakage currents (Figure 4). An additional higher-k layer may be considered to create a bilayer structure. This could reduce the field on the Al$_2$O$_3$ without significant reduction of the capacitance[56]. Altogether, our data provides design guidelines for an oxide FET, and particularly for engineering its 2DEG properties, towards achieving low off-state currents and maximizing their $I_{on}/I_{off}$ ratios.

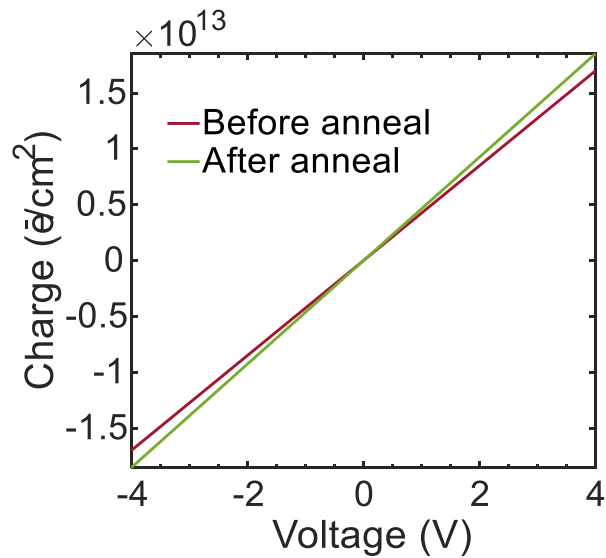

**Figure 4.** Q=CV analysis for annealed and unannealed structures, showing the possible degree of charge modulation using the studied MIS structures.

### IV. Conclusions

We show that while ALD Al$_2$O$_3$ is a mature, well-established process for gate insulator applications, optimization of small process details can lead to huge benefits in mitigating leakage. The optimized structures can be useful as gate stacks for oxide electronics, owing to their low leakage which enables significant charge modulation of an underlying channel.

Analysis of the conduction mechanisms revealed that while all those considered provide good mathematical agreement to the data, careful consideration of the extracted parameters rules out TE and FNT. Moreover, we show that the effect of the assumed flat-band voltage on the extracted parameters can be huge, necessitating careful handling of this parameter. Altogether, we identify trap-based conduction (TAT or PF) as the most likely leakage mechanism for unannealed samples. These traps can be ascribed to oxygen vacancies in $Al_2O_3$. A moderate anneal reduces the leakage currents and alleviates their temperature dependence. This observation highlights Fowler-Nordheim tunneling (FNT) as the most likely candidate, indicating the possible mitigation of the oxygen vacancies during anneal.

**Acknowledgements**

The authors are grateful for the support of the Israeli Science Foundation (ISF Grant 375/17). Partial support in the fabrication and characterization of the samples was provided by the Technion's Micro-Nano Fabrication Unit (MNFU) and the Russell Berrie Nanotechnology Institute (RBNI). We thank Prof. Felix Palumbo for fruitful discussions. Valentina Korchnoy and Arkadi Gavrilov are acknowledged for valuable technical assistance. We acknowledge the contribution of Aviv Haim and Haim Ben-Simhon, whose undergraduate research project inspired parts of this research.

**References**

[1] Y. Etinger-Geller, A. Katsman, and B. Pokroy, Chem. Mater. **29**, 4912 (2017).

[2] L. Bloch, Y. Kauffmann, and B. Pokroy, Cryst. Growth Des. **14**, 3983 (2014).

[3] M. Groner, J. Elam, F. Fabreguette, and S. George, Thin Solid Films **413**, 186 (2002).

[4] J. Robertson and R.M. Wallace, Mater. Sci. Eng. R Reports **88**, 1 (2015).

[5] M. Bosman, Y. Zhang, C.K. Cheng, X. Li, X. Wu, K.L. Pey, C.T. Lin, Y.W. Chen, S.H. Hsu, and C.H. Hsu, Appl. Phys. Lett. **97**, 103504 (2010).

[6] L. Kornblum, B. Meyler, C. Cytermann, S. Yofis, J. Salzman, and M. Eizenberg, Appl. Phys. Lett. **100**, 62907 (2012).

[7] S. Fadida, M. Eizenberg, L. Nyns, S. Van Elshocht, and M. Caymax, Microelectron. Eng. **88**, 1557 (2011).

[8] T. Ando, P. Hashemi, J. Bruley, J. Rozen, Y. Ogawa, S. Koswatta, K.K. Chan, E.A. Cartier, R. Mo, and V. Narayanan, IEEE Electron Device Lett. **38**, 303 (2017).

[9] I. Krylov, A. Gavrilov, M. Eizenberg, and D. Ritter, Appl. Phys. Lett. **103**, 53502 (2013).

[10] J. Jang, H.-H. Choi, S.H. Paik, J.K. Kim, S. Chung, and J.H. Park, Adv. Electron. Mater. **4**, 1800355 (2018).

[11] M. Tian, H. Zhong, L. Li, and Z. Wang, J. Appl. Phys. **124**, 244104 (2018).

[12] J.H. Lee, Y.C. Lin, B.H. Chen, and C.Y. Tsai, in *2010 10th IEEE Int. Conf. Solid-State Integr. Circuit Technol.* (2010), pp. 1024–1026.

[13] S.J. Pearton, F. Ren, M. Tadjer, and J. Kim, J. Appl. Phys. **124**, 220901 (2018).

[14] T.-H. Hung, K. Sasaki, A. Kuramata, D.N. Nath, P. Sung Park, C. Polchinski, and S. Rajan, Appl. Phys. Lett. **104**, 162106 (2014).

[15] M.A. Bhuiyan, H. Zhou, R. Jiang, E.X. Zhang, D.M. Fleetwood, P.D. Ye, and T.-P. Ma, IEEE Electron Device Lett. **39**, 1022 (2018).

[16] N. Pryds and V. Esposito, J. Electroceramics **38**, 1 (2017).

[17] J. Mannhart, D.H.A. Blank, H.Y. Hwang, A.J. Millis, and J.-M. Triscone, MRS Bull. **33**, 1027 (2008).

[18] L. Kornblum, Adv. Mater. Interfaces 1900480 (2019).

[19] C. Woltmann, T. Harada, H. Boschker, V. Srot, P.A. Van Aken, H. Klauk, and J. Mannhart, Phys. Rev. Appl. **4**, 1 (2015).

[20] M. Hosoda, Y. Hikita, H.Y. Hwang, and C. Bell, Appl. Phys. Lett. **103**, 103507 (2013).

[21] B. Förg, C. Richter, and J. Mannhart, Appl. Phys. Lett. **100**, 53506 (2012).

[22] Y. Chen, N. Pryds, J.E. Kleibeuker, G. Koster, J. Sun, E. Stamate, B. Shen, G. Rijnders, and S. Linderoth, Nano Lett. **11**, 3774 (2011).

[23] S.W. Lee, Y. Liu, J. Heo, and R.G. Gordon, Nano Lett. **12**, 4775 (2012).

[24] H.J. Lee, T. Moon, C.H. An, and C.S. Hwang, Adv. Electron. Mater. **5**, 1800527 (2019).

[25] T.J. Seok, Y. Liu, H.J. Jung, S. Bin Kim, D.H. Kim, S.M. Kim, J.H. Jang, D.-Y. Cho, S.W. Lee, and T.J. Park, ACS Nano **12**, 10403 (2018).

[26] T. Moon, H.J. Jung, Y.J. Kim, M.H. Park, H.J. Kim, K. Do Kim, Y.H. Lee, S.D. Hyun, H.W. Park, S.W. Lee, and C.S. Hwang, APL Mater. **5**, 42301 (2017).

[27] S.M. Kim, H.J. Kim, H.J. Jung, J.-Y. Park, T.J. Seok, Y.-H. Choa, T.J. Park, and S.W. Lee, Adv. Funct. Mater. **0**, 1807760 (n.d.).

[28] A.M. Kamerbeek, R. Ruiter, and T. Banerjee, Sci. Rep. **8**, 1378 (2018).

[29] A.M. Kamerbeek, E.K. de Vries, A. Dankert, S.P. Dash, B.J. van Wees, and T. Banerjee, Appl. Phys. Lett. **104**, 212106 (2014).

[30] D. Cohen-Azarzar, M. Baskin, and L. Kornblum, J. Appl. Phys. **123**, 245307 (2018).

[31] Z. Guo, F. Ambrosio, and A. Pasquarello, Appl. Phys. Lett. **109**, 62903 (2016).

[32] D. Liu, S.J. Clark, and J. Robertson, Appl. Phys. Lett. **96**, 32905 (2010).

[33] R.L. Puurunen, J. Appl. Phys. **97**, 121301 (2005).

[34] T.-H. Hung, K. Sasaki, A. Kuramata, D.N. Nath, P. Sung Park, C. Polchinski, and S. Rajan, Appl. Phys. Lett. **104**, 162106 (2014).

[35] H. Spahr, S. Montzka, J. Reinker, F. Hirschberg, W. Kowalsky, and H.-H. Johannes, J. Appl. Phys. **114**, 183714 (2013).

[36] J. Zhang, D. Doutt, T. Merz, J. Chakhalian, M. Kareev, J. Liu, and L.J. Brillson, Appl. Phys. Lett. **94**, 92904 (2009).

[37] Y. Taur and T.H. Ning, *Fundamentals of Modern VLSI Devices* (Cambridge university press, 2013).

[38] T. Moon, H.J. Lee, K. Do Kim, Y.H. Lee, S.D. Hyun, H.W. Park, Y. Bin Lee, B.S. Kim, and C.S. Hwang, Adv. Electron. Mater. **4**, 1800388 (2018).

[39] L. Kornblum, J.A. Rothschild, Y. Kauffmann, R. Brener, and M. Eizenberg, Phys. Rev. B **84**, 155317 (2011).

[40] J. Robertson, J. Vac. Sci. Technol. B Microelectron. Nanom. Struct. **18**, 1785 (2000).

[41] E.H. Nicollian, J.R. Brews, and E.H. Nicollian, *MOS (Metal Oxide Semiconductor) Physics and Technology* (Wiley New York et al., 1982).

[42] I. Krylov, B. Pokroy, M. Eizenberg, and D. Ritter, J. Appl. Phys. **120**, 124505 (2016).

[43] Y. Etinger-Geller, E. Zoubenko, M. Baskin, L. Kornblum, and B. Pokroy, J. Appl. Phys. **125**, 185302 (2019).

[44] S.M. Sze and K.K. Ng, *Physics of Semiconductor Devices* (Wiley, 2006).

[45] D.S. Jeong, H.B. Park, and C.S. Hwang, Appl. Phys. Lett. **86**, 72903 (2005).

[46] P. Vitanov, A. Harizanova, T. Ivanova, and T. Dimitrova, Thin Solid Films **517**, 6327 (2009).

[47] P. Kumar, M. Wiedmann, C. Winter, and I. Avrutsky, Appl. Opt. **48**, 5407 (2009).

[48] H. Schroeder, J. Appl. Phys. **117**, 215103 (2015).

[49] S. Fleischer, P.T. Lai, and Y.C. Cheng, J. Appl. Phys. **73**, 3348 (1993).

[50] S. Fleischer, P.T. Lai, and Y.C. Cheng, J. Appl. Phys. **72**, 5711 (1992).

[51] O.A. Dicks, J. Cottom, A.L. Shluger, and V. V Afanas'ev, Nanotechnology **30**, 205201 (2019).

[52] S. Yu, X. Guan, and H.-S.P. Wong, Appl. Phys. Lett. **99**, 63507 (2011).

[53] A. Gehring, *Simulation of Tunneling in Semiconductor Devices* (2003).

[54] D.S. Jeong and C.S. Hwang, J. Appl. Phys. **98**, 113701 (2005).

[55] M. Boucherit, O. Shoron, C.A. Jackson, T.A. Cain, M.L.C. Buffon, C. Polchinski, S. Stemmer, and S. Rajan, Appl. Phys. Lett. **104**, 182904 (2014).

[56] N. Kumar, A. Kitoh, and I.H. Inoue, Sci. Rep. **6**, 25789 (2016).